\documentclass{aa}  

\usepackage{graphicx}
\usepackage{txfonts}
\usepackage{wasysym}
\usepackage{enumitem}

\usepackage{threeparttable}
\usepackage{lipsum}
\usepackage{subcaption}         
\usepackage{lscape}             
\usepackage{placeins}           
\usepackage[colorlinks,citecolor=blue,urlcolor=blue,filecolor=blue,linkcolor=blue]{hyperref}
\usepackage{txfonts}
\usepackage{siunitx}

\addto\extrasenglish{
    \def\sectionautorefname{Sect.}
}
\addto\extrasenglish{
    \def\subsectionautorefname{Sect.}
}
\addto\extrasenglish{
    \def\figureautorefname{Fig.}
} 
\addto\extrasenglish{
    \def\equationautorefname{Eq.}
}

\newcommand{\Autoref}[1]{%
  \begingroup
  \renewcommand{\figureautorefname}{Figure}%
  \renewcommand{\sectionautorefname}{Section}%
  \renewcommand{\subsectionautorefname}{Section}%
  \renewcommand{\equationautorefname}{Equation}%
  \autoref{#1}%
  \endgroup}

\begin{document}

\title{Coherent radio emission at 45\,MHz from the nearby planet-hosting star GJ\,725A}

   \author{H.~W.~Edler\inst{1}\thanks{Corresponding author: edler@astron.nl} 
        \and J.~R.~Callingham\inst{1,2}
        \and I.~Davis\inst{1}
        \and C.~Cordun\inst{1,3}
        \and F.~de~Gasperin\inst{4}
        \and H.~L.~Le\inst{5,6,7}
        \and B.~J.~S.~Pope\inst{8}
        \and H.~K.~Vedantham\inst{1,3}
        \and R.~J.~van Weeren\inst{9}
        \and P.~Zarka\inst{10,11}
        }
       \institute{ASTRON, Netherlands Institute for Radio Astronomy, Oude
    Hoogeveensedijk 4, 7991 PD Dwingeloo, The Netherlands
\and
Anton Pannekoek Institute for Astronomy, University of Amsterdam, Science Park 904, 1098\,XH, Amsterdam, The Netherlands 
\and
Kapteyn Astronomical Institute, University of Groningen, PO Box 800, 9700 AV Groningen, The Netherlands
\and
INAF - Istituto di Radioastronomia, via P. Gobetti 101, Bologna, Italy
\and 
Dipartimento di Fisica, Universit\`a degli Studi di Roma Tor Vergata, via della Ricerca Scientifica 1, 00133 Rome, Italy
\and
Universit\'e C{\^o}te d'Azur, Observatoire de la C{\^o}te d'Azur, CNRS, Laboratoire Lagrange, France
\and
Department of Astronomy, Faculty of Mathematics, University of Belgrade, Studentski trg 16, 11000 Belgrade, Serbia
\and
Astronomy and Space Technology Research Centre, School of Mathematical \& Physical Sciences, Macquarie University, Sydney, NSW 2109, Australia
\and
Leiden Observatory, Leiden University, PO Box 9513, NL-2300 RA Leiden, The Netherlands
\and
LIRA, Observatoire de Paris, CNRS, Universit\'e PSL, Sorbonne Univ., Univ. Paris Cit\'e, 92190 Meudon, France
\and
ORN, Observatoire Radioastronomique de Nan\c{c}ay, Observatoire de Paris, CNRS, Univ. PSL, Univ. Orl\'eans, 18330 Nan\c{c}ay,
France
}

   \date{Received 14 August 2026; accepted 05 November 2026}

  \abstract{
  Ultralow-frequency ($\nu<100\,$MHz) coherent radio emission from exoplanet-hosting stellar systems carries information about planetary and stellar magnetic fields as well as the space weather environment -- information that is difficult to obtain at other wavelengths. We  carried out the circularly polarized LOFAR LBA Sky Survey, the most sensitive circularly polarized survey in this frequency range, using the LOFAR low-band antenna system. As part of this untargeted search, we detected a signal in a 1\,h observation that is coincident with the quiescent, planet-hosting, solar-neighborhood M\,dwarf GJ\,725A at a distance of 3.5\,pc. The signal has a high circular polarization fraction of $74\pm16\%$ and a peak brightness of $B_{I,46}=135\pm17\,\mathrm{mJy\,beam^{-1}}$ between 42\,MHz and 50\,MHz. It was undetected in the 50--66\,MHz range, with a 3$\sigma$ upper limit of $B_{I,58}<48\,\mathrm{mJy\,beam^{-1}}$, corresponding to a drop of at least a factor of ${\sim}3$ across 12\,MHz and indicating a strong spectral feature or cutoff around 50\,MHz. Our detection represents the lowest-frequency confirmed detection of a stellar system.
  Based on its high estimated brightness temperature (${\sim}10^{14}\,$K), transient character, and polarization fraction, the emission is most plausibly attributed to the electron cyclotron maser instability. 
  The available observations are not sufficient to identify the astrophysical origin of the emission. We consider three plausible scenarios: auroral emission of a yet undetected wide-orbit giant planet with a magnetic field $B\ge18\,$G, star-planet interaction with the known 11\,d orbit terrestrial planet GJ\,725Ab, or an origin due to stellar magnetic activity. Ultralow-frequency monitoring, combined with new, preferably simultaneous observations in the X-ray, UV, and H$\alpha$ and a search for a wide-orbit (thousands of days) giant planet across the full, permitted parameter space will allow us to distinguish between these scenarios.}
  
   {}
   {}
   {}
   {}
   {}

   \maketitle
   \nolinenumbers

\nolinenumbers

\section{Introduction}

Stellar activity is a central factor controlling the atmospheric evolution of planets and, ultimately, their habitability. Flares, mass ejections, and winds can erode planetary atmospheres \citep[e.g.,][]{2007AsBio...7..185L,2020IJAsB..19..136A}. This is of particular concern for planets around M dwarfs -- common hosts of temperate rocky planets in the solar neighborhood -- because their habitable zones lie close to the host stars \citep{1993Icar..101..108K,2013ApJ...765..131K}. A sufficiently strong planetary magnetic field may shield the atmosphere from this influence \citep{2007AsBio...7..167K,2013A&A...557A..67V}. Understanding the stellar activity of M dwarfs and the magnetic fields of their exoplanets is therefore central to assessing the habitability of planetary systems.

Constraining exoplanetary magnetic fields observationally is challenging. Arguably, the most promising avenue for their detection is via radio emission generated by the electron cyclotron maser instability (ECMI) that is powered or mediated by a planet. 
One possibility is that ECMI originates directly in the planetary magnetosphere, in analogy to the auroral radio emission of magnetized Solar System planets \citep{1998JGR...10320159Z}. Because ECMI radiation is emitted close to the local electron cyclotron frequency, $\nu_\mathrm{c}\approx2.8{B}$ (with $\nu_\mathrm{c}$ in megahertz and $B$ in gauss), its spectrum directly measures the magnetic field strength $B$ in the planetary magnetosphere. Alternatively, ECMI can arise in the stellar magnetosphere, powered by a sub-Alfv\'enic star-planet interaction (SPI), a scaled analog of the Jupiter-Io system \citep{2007P&SS...55..598Z}. Since the energetics of this interaction depend on the magnetic properties of the planet, SPI detections constrain the planetary magnetic fields indirectly \citep{2024NatAs...8.1359C,2026Sci...393..403R}.

However, coherent radio emission from the stellar magnetosphere can also be powered by stellar magnetic activity and rotation alone: stellar flares can be accompanied by both plasma emission and ECMI \citep{2001A&A...374.1072S}, and the breakdown of corotation of outflowing plasma provides a further channel for ECMI in rapidly rotating stars \citep[e.g.,][]{Callingham2021}. These mechanisms, if clearly identified, in turn probe the stellar wind and magnetic field.
The different origins of coherent radio emission from planetary systems can be distinguished through the brightness temperature of the signal, its degree of circular polarization, its periodicity, and its correlation with planet ephemerides or stellar activity proxies such as X-ray and ultraviolet luminosity and chromospheric emission lines \citep{2002ARA&A..40..217G,Vedantham2021,2024NatAs...8.1359C}. 

At frequencies above 100\,MHz, systematic searches for such coherent emission from M dwarfs and exoplanet systems have matured considerably in recent years \citep[e.g.,][]{2017ApJ...836L..30L,2021NatAs...5.1233C,konijn25,2026NatAs..10..410T}.
Magnetic SPI has been proposed as an explanation for the coherent radio emission from the M\,dwarfs GJ\,1151 \citep{Vedantham2020}, Proxima Centauri \citep{pereztorres2021, 2026A&A...711A.116Z}, and YZ Ceti \citep{pineda2023}. However, these SPI claims remain unconfirmed, as the star has, in some cases, not been redetected, while in others no radio periodicity or predicted close-in planets consistent with SPI could be confirmed \citep[e.g.,][]{2021A&A...649L..12P}.

Searches at ultralow frequencies ($\nu{<}100\,$MHz) are of particular interest because only this regime permits the direct detection of ECMI from the magnetospheres of Jupiter-like planets based on known dynamo laws \citep[e.g.,][]{2009Natur.457..167C}. For example, the decametric emission of Jupiter itself, with its maximum surface field of ${\sim}20\,$G, cuts off near 40\,MHz \citep{1998JGR...10320159Z,2017A&A...604A..17M}.
However, despite sustained efforts, no clear detection of coherent ultralow-frequency emission from an M dwarf or exoplanet system has been achieved. Early claims of decametric bursts from nearby flare stars with the Ukrainian T-shaped radio telescope remained unconfirmed \citep[e.g.,][]{1998A&AT...17..221A}. \citet{2021A&A...645A..59T} report a tentative signal from $\tau$~Boo\,b using beam-formed observations with the Low-Frequency Array low-band antenna \citep[LOFAR LBA;][]{2013A&A...556A...2V}. Nevertheless, follow-up campaigns have not confirmed this signal \citep{2023pre9.conf04048T,2024A&A...688A..66T,2025A&A...693A.162C}. Additional searches with the LBA, including in the first data releases of the LOFAR LBA Sky Survey (LoLSS) and the LOFAR Decametre Sky Survey (LoDeSS), have not produced any detections \citep{2020A&A...644A.157D,deGasperin2023LOFARLBASky,2026arXiv261004698C}. 
\citet{2020MNRAS.494.4848D} report a marginal detection of EQ~Peg using observations with the Long Wavelength Array.

Most recently, a tentative signal from the vicinity of the exoplanet-hosting system HD\,189733 was reported at ${\sim}50\,$MHz with NenuFAR \citep{Zhang2025,Turner2026}. While this is among the most promising candidates at these frequencies, the signal was only detected in circular polarization and not in total intensity, and the limited angular resolution of the instrument made it difficult to establish a clear association with an optical counterpart.

In this paper, we report the detection of a highly circularly polarized, transient radio emission in the 42--50\,MHz range coincident with the nearby M dwarf GJ~725A (HD\,173739, Struve\,2398\,A), discovered in V-LoLSS (Edler et al., in prep.), an untargeted search for circularly polarized sources in LoLSS. GJ~725 is a binary system in the solar neighborhood at a distance of $d=3.53\,$pc \citep{GaiaDR32021}. The primary, GJ~725A, is an M3.0\,V-type star with a radius $R_\star = 0.35\,\mathrm{R}_\odot$ and a mass $M_\star = 0.33\,\mathrm{M}_\odot$ \citep{Cortes-Zuleta2025}. It exhibits low activity, with a rotation period of 103\,d and a weak disk-averaged mean longitudinal magnetic field $B_\ell \approx-7.6$\,G and a temporal variation between epochs of ${\pm}9.2\,$G \citep{Donati2023}.
Because opposite polarities cancel in the disk-averaged $B_\ell$, the mean unsigned surface field $\langle B \rangle$, which also includes the small-scale field, may be considerably stronger. However, the literature shows some disagreement likely due to systematic effects, with a value $\langle B \rangle = 490\pm70$\,G reported by \citet{Cristofari2023} compared to the 2$\sigma$ upper limit $\langle B \rangle < 240$\,G reported by \citet{2022A&A...662A..41R}.
The star hosts a super-Earth detected with the radial velocity (RV) method, with an orbital period $P_\mathrm{orb} \approx 11.2$\,d and a minimum mass $M_{\rm p}\sin i = 2.8 \pm 0.4\,\mathrm{M}_\oplus$  \citep{Cortes-Zuleta2025}. GJ~725A forms a wide binary with GJ~725B (HD\,173740), another slowly rotating M3.5\,V-type M dwarf, at an angular separation $\approx11''$ in 2026 with a semimajor axis $a=63\pm1\,$AU \citep{Cortes-Zuleta2025}. GJ\,725B itself hosts a confirmed 37.9 d planet \citep{Ould-Elhkim26}.

\section{LOFAR observations}\label{sec:observations}
In our untargeted V-LoLSS survey covering the 42--66\,MHz range, we detected significant emission coincident with GJ\,725 in the pointing P280+58 at $1.75^\circ$ from the pointing center (corresponding to a primary beam response level of 55\%). The source was initially discovered in a combined $1'$-resolution image created from three 1 h observations taken at different epochs across 27 days (IDs 801916, 800238, and 800728), with a peak surface brightness $B_{I,54} = 40\pm8$\,mJy\,beam$^{-1}$ and $B_{V,54} = 29\pm6$\,mJy\,beam$^{-1}$ in Stokes $I$ and $V$, respectively, which corresponds to a statistical significance $\approx5\sigma$ in both cases.

\begin{figure*}
    \centering
    \includegraphics[width=0.99\linewidth]{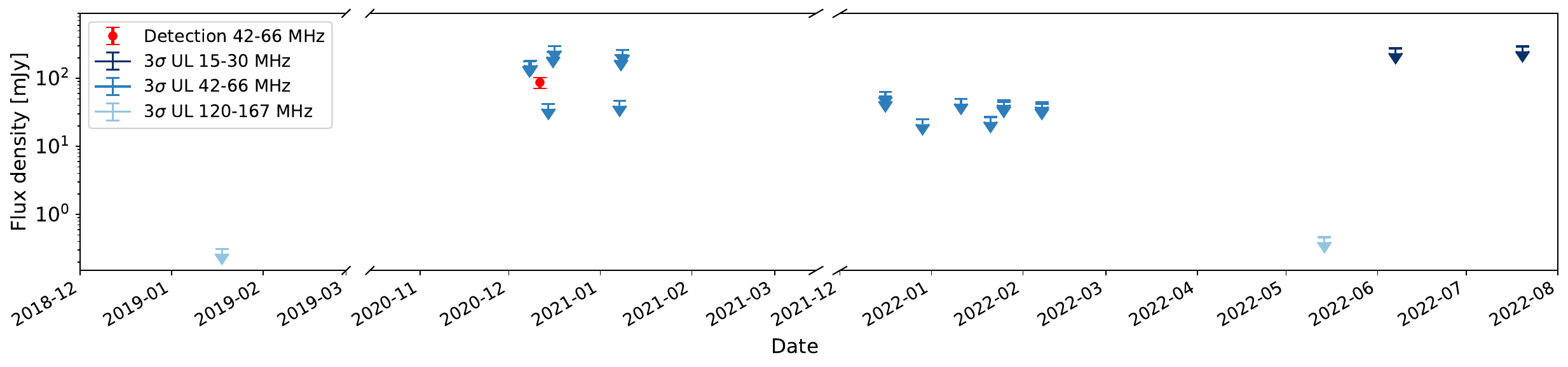}
    \caption{Date of the radio detection and 3$\sigma$ upper limits (UL). The periods  March 2019 to  October 2020 and  April 2021 to November 2021 are not shown as they contain no observations. The upper limits are color-coded according to frequency, from dark blue (lowest) to light blue (highest).}
    \label{fig:radio_time}
\end{figure*}
We imaged each LoLSS observation covering the source individually and find the emission to be transient. It was significantly detected only during the 11 December 2020 observation (ID: 801916), with a peak surface brightness $B_{I,54} {=} 87\pm16$\,mJy\,beam$^{-1}$. Notably, the source was undetected during an observation conducted three days later (ID: 800238), with a 3$\sigma$ upper limit $B_{I,54} {\leq} 42\,\mathrm{mJy\,beam^{-1}}$, confirming its transient nature.
The 1 h LoLSS observations covering the source have variable sensitivities due to their distance from GJ\,725. No significant emission is detected in any other LoLSS observation, with 11 observations exhibiting 3$\sigma$ upper limits below the flux density in the detection epoch. We further find no significant emission in total or circularly polarized intensity in the two 8\,h observations of the LOFAR Two-metre Sky Survey \citep[LoTSS;][]{2026A&A...707A.198S}.
We also processed the good-quality observations of the LoDeSS survey that cover GJ\,725 in the 15--30\,MHz range, again yielding no detection.  
In \autoref{fig:radio_time}, we display the flux density and observation date of our detection and the upper limits of the LoLSS, LoDeSS, and LoTSS observations covering the source.

We present our analysis of the epoch with the detection in greater detail, applying the standard extraction and recalibration procedure of \citet{vanWeeren2021LOFARObservationsGalaxy}. We subtracted all sources outside a 10 arc-minute circle containing the target from the visibility data and performed self-calibration to improve the image fidelity. This produced a small dataset for reimaging.

\begin{figure}
    \centering
    \includegraphics[width=0.75\linewidth]{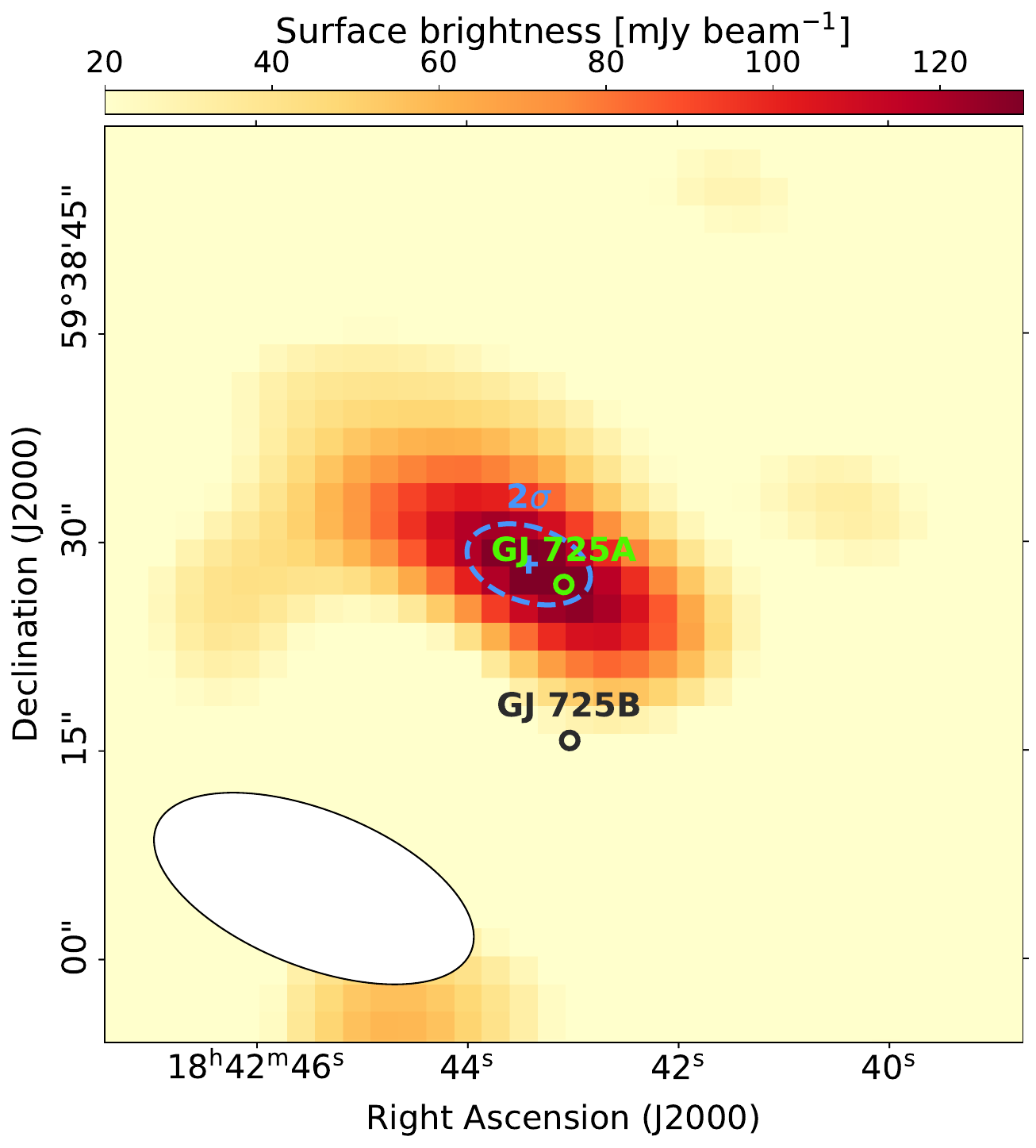}
    \includegraphics[width=0.75\linewidth]{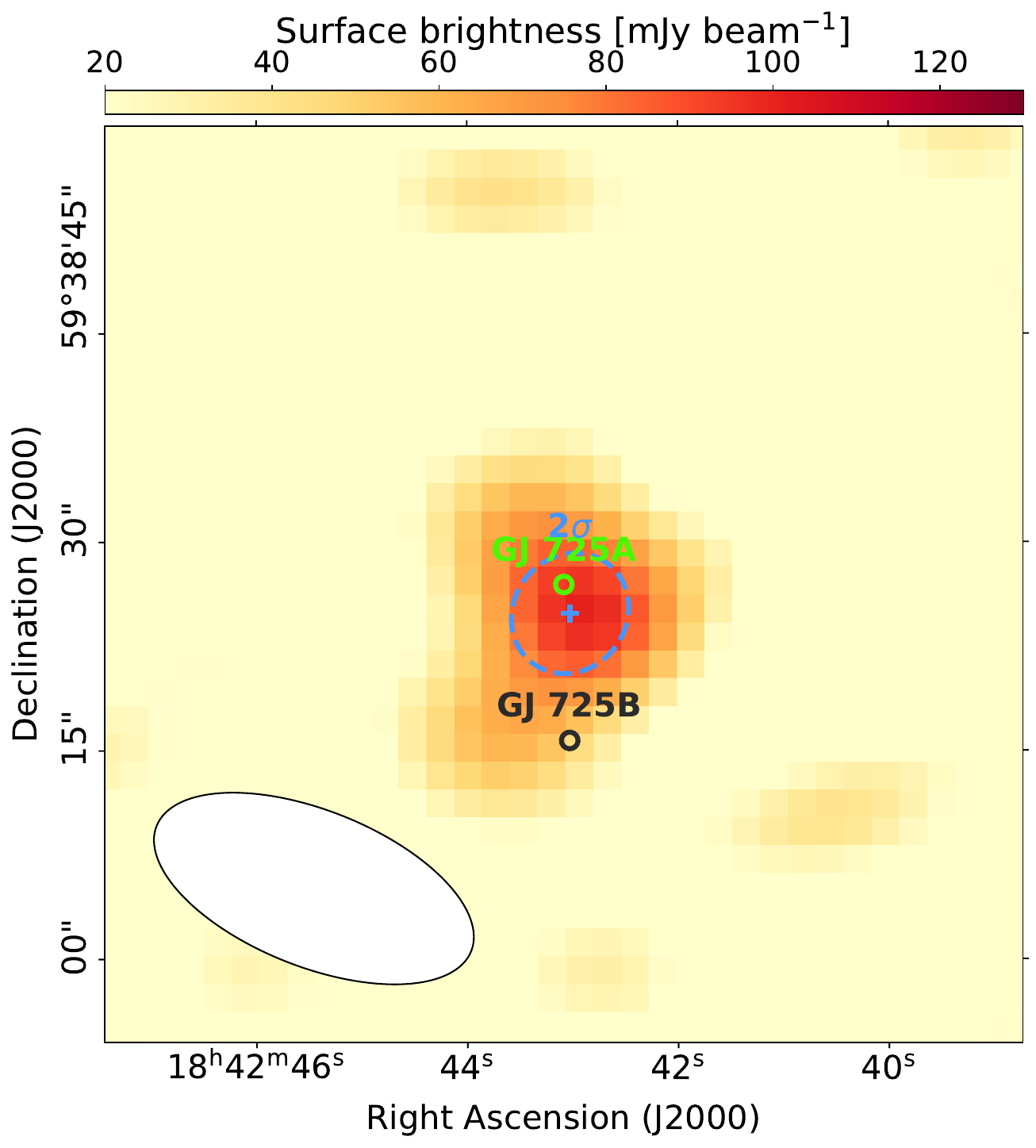}
    \caption{Total (top) and circularly polarized intensity (bottom) of the transient source at 46\,MHz, showing the $2\sigma$ localization of the radio source (blue) and the position of GJ\,725A (green) and GJ\,725B (black) at the epoch of the LBA observation. The white ellipse in the lower left indicates the synthesized beam size of $24''\times11''$. }
    \label{fig:astrometry}
\end{figure}

We find that in the 1\,h detection epoch, the source is detected only in the lower third of the band, in the frequency range $42\leq\nu\leq50\,$MHz, where the peak intensity is $B_{I,46} = 135\pm17$\,mJy\,beam$^{-1}$ and $B_{V,46} = 100\pm17$\,mJy\,beam$^{-1}$ at statistical significances of $7.9\sigma$ and $5.9\sigma$, respectively. The emission exhibits a left-hand circular polarization fraction of ${V}/{I}=74\pm16\%$. The LoLSS data follow the Stokes $V$ sign convention used in the pulsar literature: a positive flux density corresponds to left-hand circularly polarized emission, rather than following the IAU definition. The single-epoch Stokes $I$ and $V$ images of the source at 46\,MHz are displayed in \autoref{fig:astrometry}.
In the upper two-thirds of the band, no emission is detected, with a $3\sigma$ upper limit $B_{I,58} \leq48$\,mJy\,beam$^{-1}$.

\begin{figure}
    \centering
    \includegraphics[width=0.8\linewidth]{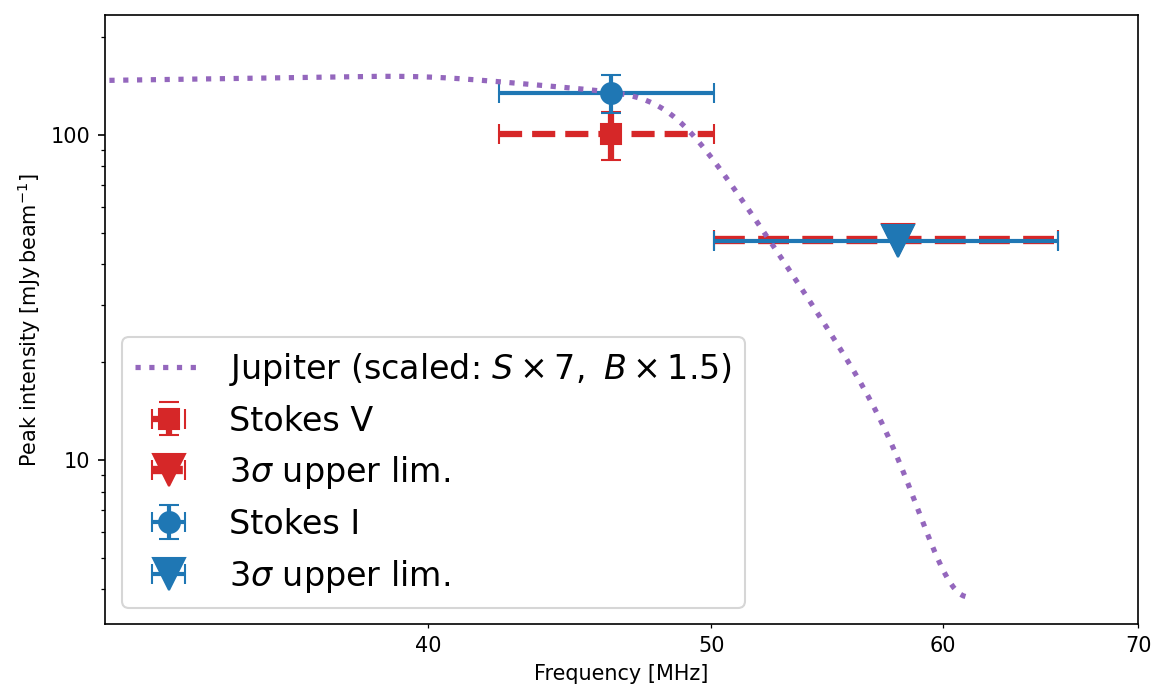}
    \caption{Peak intensity and upper limit in Stokes $I$ (blue) and $V$ (red) of GJ\,725 during the 1 h observation on 11 December 2020 (ID: 801916). As reference, we show the burst spectrum of Jupiter \citep{1998JGR...10320159Z}, scaled to a distance of 3.53\,pc by a factor of 1.5 in magnetic field strength and a factor of 7 in brightness (purple).}
    \label{fig:spectrum}
\end{figure}

\Autoref{fig:spectrum} shows the measured peak intensity together with this limit. 
At $5.9\sigma$ significance, we can exclude that the emission measured in the lower third continues into the upper two thirds of the band with the same intensity. The intensity must therefore decrease by at least a factor of $135/48\approx2.8$ within a few MHz, consistent with a spectral cutoff or equally sharp steepening near $50\,$MHz.
Our most sensitive upper limits, in the 120--168\,MHz range, therefore, do not necessarily imply that the source was inactive. Rather, they may simply reflect its intrinsic spectrum.

\subsection{Dynamic spectra}
\begin{figure}
    \centering
        \includegraphics[width=1\linewidth]{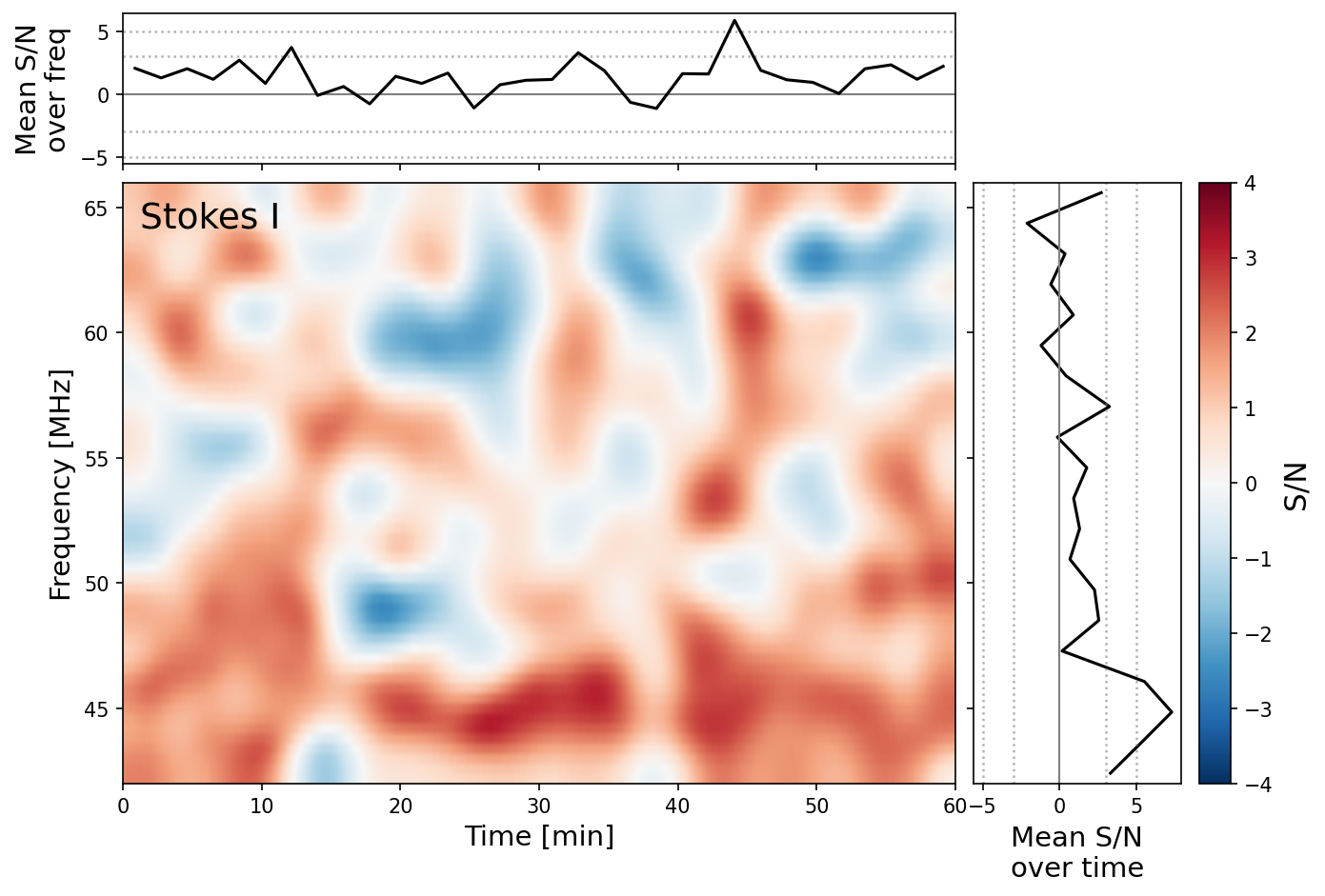}
        \includegraphics[width=1\linewidth]{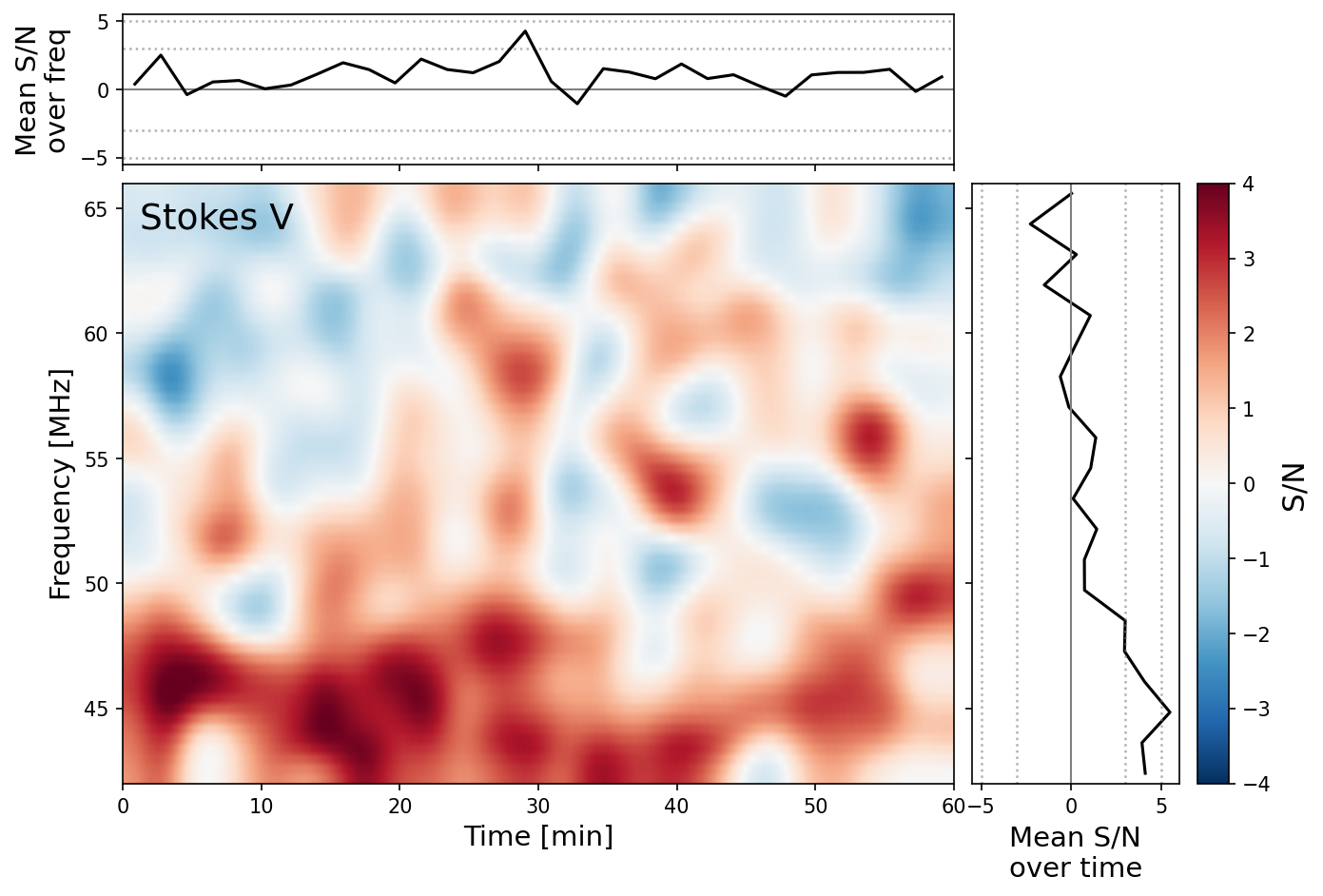}
    \caption{Dynamic spectra of GJ\,725 expressed in terms of S/N in Stokes $I$ (top) and $V$ (bottom). The dynamic spectra were convolved by a Gaussian kernel with a width of 1.17\,MHz and 112\,s. Near the top and right-hand side of the dynamic spectra, the frequency and time-collapsed significances are shown in bins corresponding to the width of the convolution kernels.}
    \label{fig:dynspecs}
\end{figure}

To analyze the temporal and spectral behavior of the detected emission, we created dynamic spectra of the source. We used the 54\,MHz sky model of the field and the direction-dependent calibration solutions from LoLSS to subtract all sources except for a $10'$ sized region covering GJ\,725 from the $uv$-data. We then phase-shifted the dataset toward GJ\,725 and applied the primary beam model and the calibration solutions toward the new phase center. We used \texttt{RIMS} \citep{2026NatAs..10..410T}  to create the dynamic spectra of the source and 150 off-source regions, which were randomly selected within a 0.3$^\circ$ radius, at a resolution of 8\,s in time and 0.39\,MHz in frequency. 
Based on the off-axis spectra, we estimated the background mean $\bar{S}_{i,j}$ and standard deviation $\sigma_{S_{i,j}}$ and calculated the signal-to-noise ratio (S/N) of each pixel according to $\mathrm{S/N} = (S_{i,j}-\bar{S}_{i,j})/\sigma_{S_{i,j}}$.
In \autoref{fig:dynspecs}, we show the Stokes $I$ and $V$ dynamic spectra in terms of S/N, smoothed with a Gaussian kernel of 1.17\,MHz and 112\,s. 
The signal reaches mostly marginal significance ($2$--$3\sigma$) throughout the observation, indicating that it is not a single short burst but persists across at least the observed hour. It shows clear frequency structure, with significant emission detected only below $50\,$MHz and peaking in significance at $45\,$MHz (consistent with \autoref{fig:spectrum}).
The Stokes $I$ dynamic spectrum shows a slightly lower S/N at low frequencies. This is due to the contribution of side-lobe confusion from imperfectly subtracted sources to the noise level, which is not a concern in Stokes $V$.
An alternative unsmoothed version of these plots, with bin sizes corresponding to the width of the Gaussian kernel described above, is displayed in \autoref{sec:app_dynspec}.

\subsection{Astrometric association with GJ 725A}
GJ\,725 is a binary system with an apparent separation $\approx11.2''$ at the epoch of our detection \citep[2020,][]{GaiaDR32021}. Given the astrometric accuracy of LoLSS, $\sigma_\mathrm{RA} = 1.45''$ and $\sigma_\mathrm{Dec}=1.12''$, across the full survey (de Gasperin et al., in prep.), the two components can be distinguished.
Using the source finder \texttt{PyBDSF} \citep{Mohan2015PyBDSFPythonBlob}, we located the transient in total intensity at $\alpha_\mathrm{J2000}{=}18^\mathrm{h}42^\mathrm{m}43.42^\mathrm{s}$, $\delta_\mathrm{J2000}{=}{+}59\degr38\arcmin 28.2\arcsec$, with source-finding uncertainties of $\sigma_\mathrm{RA} = 1.7''$, $\sigma_\mathrm{Dec}=1.1''$.

To quantify the astrometric accuracy of our data, we constructed a wide-field image and source catalog from the detection epoch over the 42--50\,MHz range and cross-matched it with LoTSS, which is aligned to \textit{Pan-STARRS} \citep{Flewelling} to an accuracy well below 1 arc second. Selecting isolated sources (no neighbor within $30''$) within $2.7^\circ$ of the pointing center, brighter than $10\,$mJy, detected above $5\sigma$, and with a reconstructed size ${\leq}45''$, we found 42 matches. These have a mean positional offset of $\Delta\alpha = 0.4 \pm 1.4''$ and $\Delta\delta = -0.3 \pm 0.9''$. \Autoref{fig:astrometry} shows the source location and its combined $2\sigma$ uncertainty region (from source finding and astrometry) relative to the proper-motion-corrected \textit{Gaia DR3} positions of GJ\,725A and GJ\,725B \citep{GaiaDR32021}.
The emission lies within $2\sigma$ of GJ\,725A, whereas an association with GJ\,725B is excluded at ${>}7\sigma$.

We repeated the localization in circular polarization, propagating the same source-finding and astrometric uncertainties. Since we lack the Stokes~$V$ sources to characterize the astrometric uncertainty independently, we adopted the Stokes~$I$ value and only determined the source-finding uncertainty in Stokes $V$. The Stokes~$V$ source position (lower panel of \autoref{fig:astrometry}) again agrees with GJ\,725A (offset $<1\sigma$) and excludes GJ\,725B at $>5\sigma$. We therefore conclude that the radio emission originates from GJ\,725A.

\section{Emission mechanism}\label{sec:mechanism}
Below we investigate the mechanism responsible for the radio emission, given the observational constraints derived in the previous section: the high degree of circular polarization ($74\pm16\%
$), the spectral cutoff or strong steepening at $\nu_\mathrm{max}\approx50\,$MHz, and the temporal structure, with emission persisting for at least 1\,h in the epoch of detection but absent in all others. 
A further constraint is the implied brightness temperature. At a distance of 3.53\,pc, it is
\begin{equation}
    T_\mathrm{b}=\frac{S\lambda^2}{2k_\mathrm{B}\Omega}\approx 1.4\times10^{14}{f}^{-1}\,\mathrm{K},
\end{equation} 
where $S$ is the peak flux density, $k_\mathrm{B}$ the Boltzmann constant, $\lambda$ the wavelength, $\Omega$ the solid angle of the emitting region, and $f$ its projected area as a fraction of the stellar disk of GJ\,725A. Since the emitting region likely does not exceed the entire stellar disk ($f\leq1$), this is a lower limit.
If the emission is isotropic, the emitted power in the $42{-}50$\,MHz band is $1.6\times10^{22}\,\mathrm{erg\,s^{-1}}$.
The brightness temperature exceeds the $\sim 10^{12}\,$K limit for incoherent synchrotron emission \citep{1969ApJ...155L..71K} and, together with the high circular polarization, requires a coherent emission process. 
At meter wavelengths, two such mechanisms operate in stellar systems: plasma emission and ECMI \citep[e.g.,][]{Vedantham2021,2024NatAs...8.1359C}.

\subsection{Plasma emission} \label{sec:plasma}
Plasma emission is generated by Langmuir waves at the local plasma frequency $\nu_\mathrm{p} \approx 9n_\mathrm{e}^{1/2}\,$ (with $\nu_\mathrm{p}$ in megahertz and $n_\mathrm{e}$ in $10^{6}\,\mathrm{cm}^{-3}$) or its harmonics, so the emission frequency traces the electron density at the source. Assuming the observed emission is fundamental, the cutoff at $\nu_\mathrm{max} \approx 50\,$MHz implies $n_\mathrm{e} = 3.1\times10^{7}\,\mathrm{cm^{-3}}$ at the emission site. For first-harmonic emission, the implied density is lower by a factor of 4.

Two conditions must be met for plasma emission to explain the signal. First, for the radiation to be generated and escape, the plasma frequency must exceed the local cyclotron frequency, $\nu_\mathrm{p} > \nu_\mathrm{c}$ \citep{2001A&A...374.1072S}. We modeled the stellar wind and magnetic field of GJ\,725A, as described in \autoref{sec:app_wind}. Due to the observational uncertainty in the parameters, we explored values an order of magnitude lower and higher than our estimate for the base coronal density of $n_0=2\times10^7\,\mathrm{cm^{-3}}$, and we probed values of the large-scale surface magnetic field $B_{\star,0}$ below 500\,G.
We find that at the radius where $\nu_\mathrm{p} = 50\,$MHz, the plasma emission formation condition is satisfied only for low values of $B_\mathrm{\star,0} < \mathrm{50\,G}$ and base electron densities $n_0\gtrsim 3\times10^{7}\mathrm{\,cm^{-3}}$ (see \autoref{sec:app_wind}).

Second, plasma emission must be able to reproduce the observed brightness temperature. Following \citet{2001A&A...374.1072S} and \citet{Vedantham2021}, we estimated the saturation brightness temperature of fundamental and harmonic plasma emission for the coronal temperature of GJ\,725A ($T = 1.8\times10^{6}\,$K in our model) and a hot plasma component  $T_1$ that is $10{-}40$ times hotter.
The predicted values -- $T_\mathrm{b,f}<4\times10^{9}$\,K for the fundamental and $T_\mathrm{b,h}\leq3\times10^{12}$\,K for the harmonic -- fall short of the observationally inferred lower limit of $T_\mathrm{b}\geq1.4\times10^{14}$\,K by two to five orders of magnitude.
Harmonic emission is incompatible with the observed circular polarization fraction of $74\pm16\%$ \citep{melrose1978,melrose1980}.
We conclude that plasma emission cannot account for the properties of the observed signal.

\subsection{ECMI}\label{sec:ecmi}
Emission from the ECMI occurs close to the local cyclotron frequency $\nu_\mathrm{c} \approx 2.8B$ (with $\nu_\mathrm{c}$ in megahertz and $B$ in gauss) or its low harmonics \citep[e.g.,][]{2006A&ARv..13..229T}. Typically, emission is expected to be dominated by the fundamental, as it has the highest growth rate \citep{1982AuJPh..35..447H}. Therefore, the maximum emitted frequency directly measures the highest magnetic field strength in the emission region $B_\mathrm{max} \approx 18\,$G.

For the ECMI to develop, the reverse of the plasma-emission condition must be satisfied, $\nu_\mathrm{c} > \nu_\mathrm{p}$. 
If the emission originates in the stellar magnetosphere, we can test this with our wind model. At the radius where $\nu_\mathrm{c} = 50\,$MHz, ECMI formation requires a base wind density
\begin{equation}\label{eq:escapefit}
    n_0 \lesssim 2.4\times10^8\left(\frac{B_\mathrm{\star,0}}{50\,\mathrm{G}}\right)^{2}\,\mathrm{cm^{-3}},
\end{equation}
which is satisfied for a large part of the considered parameter space (see \autoref{sec:app_wind}).
Unlike plasma emission, ECMI faces no difficulty with the observed energetics: the maser routinely reaches brightness temperatures far in excess of $10^{14}\,$K \citep[e.g.,][]{2006A&ARv..13..229T}, and it is the established mechanism behind the auroral radio emission of the magnetized Solar System planets \citep{1998JGR...10320159Z} and of coherent bursts from M dwarfs and ultracool dwarfs \citep{2015Natur.523..568H}.

The polarization, brightness temperature, and spectral cutoff of the detection are thus all consistent with fundamental ECMI from a region of $B \approx 18\,$G. We adopt ECMI as the emission mechanism and address the question of the origin of the emission and the required mildly relativistic electrons in the following section.

\section{Astrophysical origin of the ECMI emission}\label{sec:origin}
Having established ECMI as the likely emission mechanism, we now investigate the origin of the emission. Plausible ECMI emission regions are the magnetosphere of the M dwarf GJ\,725A or the magnetosphere of an orbiting planet.  For fundamental $x$-mode ECMI, the radial orientation of the magnetic field in the emitting region sets the handedness of the observed signal. Our detection shows positive flux density in Stokes $V$, which corresponds to left-handed polarization in the convention we adopted (see \autoref{sec:observations}), as also observed for Jupiter's southern pole \citep{2017A&A...604A..17M}. For an origin of the signal in the stellar magnetosphere, this agrees with the finding of \citet{Donati2023} that the large-scale disk-averaged field of GJ\,725A is predominantly poloidal and axisymmetric and points away from us. For emission from a planetary magnetosphere, by contrast, the handedness would reflect the (unknown) polarity of the planetary field.
In the following, we investigate three different scenarios: ECMI from a planetary magnetosphere, ECMI in the stellar magnetosphere powered by stellar activity, and ECMI in the stellar magnetosphere that is powered by a magnetic SPI.

\subsection{ECMI from a planetary magnetosphere}
Given the ultralow frequency of the signal, the inferred magnetic field strength of 18\,G in the emitting region falls in a range that is realistic for planetary magnetospheres and comparable to Jupiter's maximum surface field strength \citep[${\sim}20\,$G;][]{2022JGRE..12707055C}. However, given the low mass of the only confirmed planet GJ\,725Ab ($M_{\rm p}\sin i = 2.8 \pm 0.4\,\mathrm{M}_\oplus$), dynamo scaling laws predict a magnetic field at least an order of magnitude below the required strength \citep{2009Natur.457..167C}. This suggests that GJ\,725Ab is unlikely to be the emission site. An exoplanetary origin of the radio emission therefore likely requires a yet-undetected giant planet with a field at least as strong as that of Jupiter.

The parameter space for such a planet is constrained: \citet{Cortes-Zuleta2025} exclude the presence of a planet with $M\sin{i}\geq10\,\mathrm{M}_\oplus$ in an orbit below 800 days based on RV data. Moreover, the absence of a \textit{Gaia-Hipparcos} proper-motion anomaly \citep[PMA;][]{2019A&A...623A..72K} disfavors massive planets in orbits comparable to the temporal baseline of the astrometric measurements of ${\sim}9\,000$\,days. However, this orbital period interval is not fully probed due to the limited temporal observing windows and affected by the binary companion. Orbits longer than 29\,000 days would be unstable due to the binary companion \citep{Cortes-Zuleta2025}.

Furthermore, the stellar-wind magnetic power for driving planetary auroral emission is weak at the wide orbital separations considered here. Using our magnetic radiometric model (Cordun et al., in prep.) based on the magnetic form of the radiometric Bode's law \citep{2007P&SS...55..598Z,griessmeier2007predicting}, we estimate that a $1\,R_{\rm J}$, $3$--$15\,M_{\rm J}$ companion at 6\,au would produce a radio flux density approximately seven orders of magnitude below the observed signal. Our calculation assumes that a fraction of $2\times10^{-3}$ of the stellar-wind magnetic power intercepted by the planetary magnetosphere is converted into radio emission \citep{2007P&SS...55..598Z}, beamed into a solid angle of $1.6$\,sr \citep{zarka2004jupiter} over a bandwidth of $0.5\nu_{\rm c}$ \citep{1998JGR...10320159Z,griessmeier2007predicting}. Instead, the emission could be powered via the breakdown of corotation of plasma provided by a volcanically active satellite (as with Jupiter's auroral emission). According to the model of \citet{2011MNRAS.414.2125N}, this mechanism would be able to reproduce the observed flux density from the emission of a full auroral oval for a Jupiter-like planet if it rotates at ${\sim}5$ times the rate of Jupiter.
Thus, while the permitted range of orbital periods for a massive planet in the system is limited and an internal source of energetic electrons would need to be invoked, direct auroral emission from a Jupiter-like planet at a wide orbit remains an exciting, albeit speculative, scenario.

\begin{table}
\centering
\caption{Stellar properties and activity indicators of GJ\,725A,
the low-frequency SPI candidate GJ\,1151, and the prototypical
flare star AD~Leo.}
\label{tab:comparison}
\renewcommand{\arraystretch}{1.25}
\begin{tabular}{lccc}
\hline\hline
Parameter & GJ\,725A & GJ\,1151 & AD~Leo \\
\hline
Spectral type & M3.0\,V & M4.5\,V & M3.0\,V \\
Distance [pc] & 3.53 & 8.04 & 4.97 \\
$L_{\mathrm{H}\alpha}/L_{\mathrm{bol}}$ [$10^{-4}$] & 0.072 & 0.067 & 1.72 \\
$L_{\mathrm{X}}/L_{\mathrm{bol}}$ [$10^{-5}$] & $<1.4$ & $<1.07$ & 105.7 \\
$\mathit{Ro}$\tablefootmark{(a)} & 1.77 & 0.98 & 0.034 \\
$B_\ell$ (mean|variation)\tablefootmark{(b)} [G]     & $-7.6 \,|\, 9.2$ & $1.1 \,|\, 16.3$     & $-207 \,|\, 56$ \\

\hline
\end{tabular}
\tablefoot{
\tablefoottext{a}{Rossby number, computed from the rotation periods \citep{2024MNRAS.527.4330L, 2016ApJ...822...97H} and the
convective turnover time following \citet{2018MNRAS.479.2351W}.}
\tablefoottext{b}{Mean and temporal variation (in terms of standard deviation) of the longitudinal magnetic field measurements
\citep{2023A&A...676A..56B, Donati2023, 2024MNRAS.527.4330L}.}
}
\end{table}
\subsection{ECMI powered by stellar activity}\label{sec:ecmi_stellar}

Long-duration ECM bursts are commonly detected from active M dwarfs \citep[e.g.,][]{2017ApJ...836L..30L,2019ApJ...871..214V,Callingham2021,2024A&A...682A.170B,2025A&A...695A..95Z} and are typically attributed either to auroral emission, driven by the breakdown of corotation between the magnetic field and outflowing magnetospheric plasma, or to flaring coronal loops. 
Such emission is generally observed from well-known flare stars with high levels of magnetic activity, although low-frequency ECMI emission has also been detected from a small number of quiescent systems, for which SPI has been proposed as the power source (see \autoref{sec:spi}).

Auroral emission of the type observed for Jupiter and for ultracool dwarfs \citep{1998JGR...10320159Z,2017MNRAS.470.4274T} is powered by rotation, with periods of hours to days. For the slow rotation of GJ\,725A (103\,d), this mechanism is orders of magnitude less efficient and cannot plausibly account for the observed signal \citep{2017MNRAS.470.4274T}.

Emission from flaring coronal loops would instead be expected to correlate with indicators of stellar activity.
One such indicator is the Rossby number, the ratio of rotation period to convective turnover time, which for GJ\,725A is $\mathit{Ro}\approx1.77$, placing it firmly in the unsaturated regime of inefficient dynamo action, consistent with an old ($\sim6\,$Gyr), slowly rotating, and low-activity star \citep{Cortes-Zuleta2025}.
Indeed, no M dwarf previously detected by LOFAR has a higher Rossby number. The same picture emerges from other activity indicators, such as the H$\alpha$ luminosity and the longitudinal magnetic field. In \autoref{tab:comparison}, we compare GJ\,725A with the quiescent SPI candidate GJ\,1151 and the radio-loud flare star AD Leo.
We further estimated an upper limit of GJ\,725's optical flare rate from TESS  \citep[see \autoref{sec:app_flarerate}]{2015JATIS...1a4003R}, which covered GJ\,725 in 42 sectors from July 2019 to December 2024. While GJ\,725 was not covered by TESS during our radio detection, we find no flares two years prior or after that, with most flares occurring after April 2024. Our analysis yields a flare rate ${<}0.013\,\mathrm{d}^{-1}$, compared to rates ${\sim}1\,\mathrm{d}^{-1}$ for active, LOFAR-detected flare stars \citep{2021ApJ...919L..10P}. 

We caution, however, that no clear connection between low-frequency coherent emission and coronal activity indicators or optical and X-ray flare rates has yet been established for M dwarfs \citep{Callingham2021, 2024A&A...684A...3Y}. The quiescence of GJ\,725A therefore makes a stellar origin of the radio emission less likely but does not by itself exclude it.

\subsection{ECMI from star-planet interaction}\label{sec:spi}

\begin{figure}
    \centering
    \includegraphics[width=0.94\linewidth]{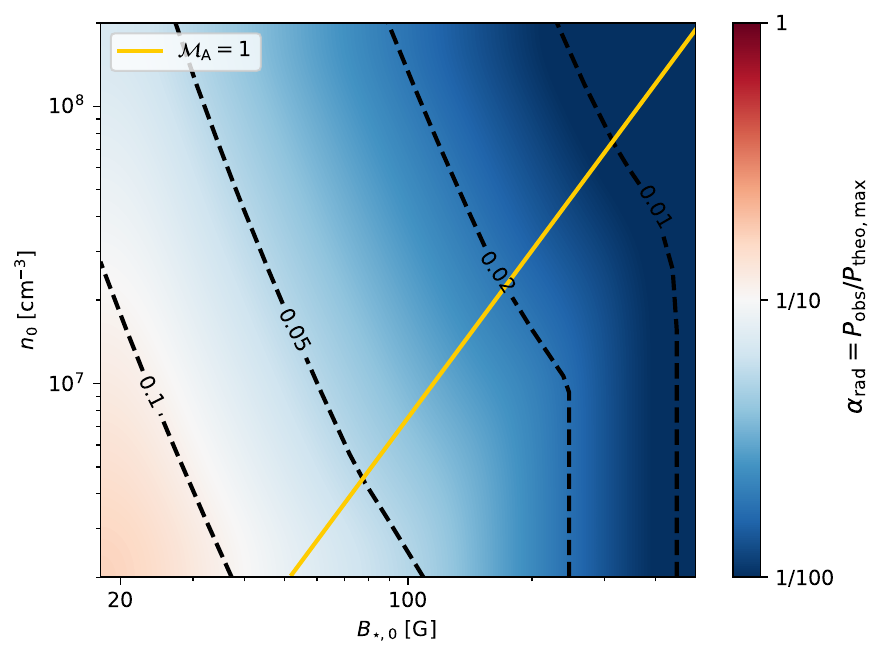}
    \caption{Efficiency ratio $\alpha_\mathrm{rad}$ between the observed power and maximum power directed toward the star in SPI, as predicted following \citet{2026A&A...705A..12P}. Values are shown for a magnetized planet with an equatorial surface field of 0.5\,G and a radius of $1.7\,\mathrm{R_\oplus}$. The area to the lower right of the yellow line is the regime in which sub-Alfv\'enic SPI can take place, and the dashed black lines mark the 10\%, 5\%, and 2\% contours. The color scale is centered on the highest proposed radiative efficiency in the literature \citep[10\%;][]{2011A&A...526A.161K}.}
    \label{fig:spi_energy}
\end{figure}

Given the difficulties of explaining the observed emission through stellar activity, it may instead be powered by an  interaction with the known close-in planet GJ\,725Ab.
For SPI to take place, the Alfv\'enic Mach number $\mathcal{M}_\mathrm{A}=v_\mathrm{rel}/v_\mathrm{A}$, given by the ratio of the wind velocity in the planet's rest frame ($v_\mathrm{rel}$) to the local Alfv\'en velocity ($v_\mathrm{A}$), must be $<1$. 
Based on our model of the stellar wind and magnetic field, we investigate the part of the parameter space that satisfies this condition. The part of the parameter space where the orbit is sub-Alfv\'enic is displayed in \autoref{fig:spi_energy} (see also \autoref{sec:app_wind}) and can be parameterized as
\begin{equation}\label{eq:subalfven}
    n_{0} \lesssim 1.9\times10^6\left(\frac{B_\mathrm{\star,0}}{50\,\mathrm{G}}\right)^{2}\,\mathrm{cm^{-3}}.
\end{equation}
Thus, the planet is sub-Alfv\'enic only for a tenuous wind and/or a strong large-scale magnetic field. 
For all values where the planet is sub-Alfv\'enic, the condition for ECMI emission  $\nu_\mathrm{c} > \nu_\mathrm{p}$ is also fulfilled.

\begin{figure*}
    \centering
        \includegraphics[width=0.99\linewidth]{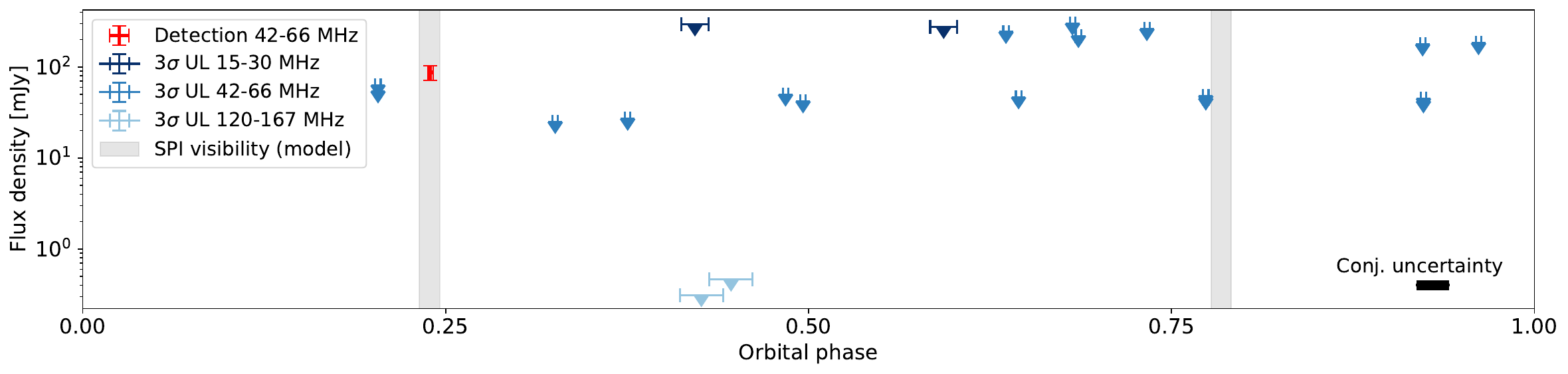}
        \vspace{-2mm}
    \caption{Planetary orbital phase of the radio detection and upper limits. The extent of the points along the $x$-axis corresponds to the duration of the observations and not to the uncertainty. The uncertainty on the orbital phase (${\geq}0.045$) is instead indicated by the black bar in the bottom. The gray-shaded regions indicate the maser visibility windows for a model of a rotation-aligned dipole that emits at the detected phase. For display purposes, the model curves include a minor shift $\Delta\phi=0.004$, well within the ephemeris uncertainty (see \autoref{sec:app_maser}).}
    \label{fig:radio_period}
\end{figure*}

An important check for the feasibility of SPI is whether the power of the sub-Alfv\'enic interaction is sufficient to explain the observed radio emission,
\begin{equation}
    P_\mathrm{obs}=\Omega_\mathrm{rad} d^2 \Delta\nu S\approx2.0\times10^{20}\,\mathrm{erg\,s^{-1}},
\end{equation}
where $\Omega_\mathrm{rad}\approx 0.16\,$sr is the solid angle of the ECMI radiation beam of a single flux tube \citep{2000JGR...10516053K}, $\Delta\nu\approx8\,$MHz the bandwidth across which the signal is detected, and $S$ the observed flux density.
Various models estimate the expected radio emission from SPI \citep[e.g.,][]{2007P&SS...55..598Z,2009A&A...505..339L,2011A&A...526A.161K,2013A&A...552A.119S}. These models rely on assumptions regarding how efficiently the Poynting flux intercepted by the planet is transferred to the stellar magnetosphere. They can substantially differ. \citet{2026A&A...705A..12P} recently extended these models based on 3D magnetohydrodynamic simulations and derived the following relation for the maximum possible power that an SPI with a magnetized planet can direct toward the stellar surface:
\begin{equation}\label{eq:paul}
P_{\mathrm{theo,\max}}=5.38R_{\mathrm{p}}^{2}B_{\mathrm{sw}}^{2}v_{\mathrm{rel}}\left(\frac{B_{\mathrm{p}}}{1\,\mathrm{G}}\right)^{0.5}\left(\frac{1\,\mathrm{G}}{B_{\mathrm{sw}}}\right)^{0.85}. 
\end{equation}
Here, all quantities are expressed in cgs units; $R_\mathrm{p}$ and $B_\mathrm{p}$ represent the radius and magnetic field of the planet, $v_\mathrm{rel}$ is the relative velocity between the stellar wind and the planet, and $B_\mathrm{sw}$ is the stellar wind magnetic field at the planet's orbit.
Only a fraction $\alpha_\mathrm{rad}=P_\mathrm{obs}/P_\mathrm{theo,\max}<1$ of the redirected power is converted to radio emission, this radiative efficiency is estimated to be in the range $1-10\%$ \citep{2006A&ARv..13..229T,2007P&SS...55..598Z,2011A&A...526A.161K}.
In \autoref{fig:spi_energy}, we display the ratio $\alpha_\mathrm{rad}$ between the observationally inferred power and the maximum available power from \autoref{eq:paul} for a planet with a fiducial equatorial surface magnetic field of $B_\mathrm{p}{=}0.5\,$G and a radius of $1.7\,\mathrm{R_\oplus}$ (corresponding to the larger radius of the two maxima of the bimodal probability distribution derived by \citealt{Cortes-Zuleta2025}). For these parameters, SPI is energetically viable with an efficiency of $5\%$ or less for a base strength of the large-scale field $B_{\star,0}>80\,$G.

In \autoref{fig:radio_period}, we show the radio detection and
upper limits as a function of orbital phase. For SPI, the signal
is expected to be periodic with the planetary orbit. In the
Jupiter-Io system, the emission appears near the orbital
quadratures \citep{2017A&A...604A..17M}, although
\citet{2023MNRAS.524.6267K} show that the favored phases
depend on the stellar magnetic and rotational
geometry and do not generally have to coincide with the quadrature. Our detection occurs at an orbital phase of $\phi_\mathrm{det}{=}0.24\pm0.05$, close to
quadrature and thus consistent with the Jupiter--Io analogy for
the near-axisymmetric field topology reported for GJ\,725A
\citep{Donati2023}. 
In \autoref{sec:app_maser} we explore which combinations of
stellar inclination $i_\star$ and magnetic obliquity $\beta$ are compatible with emission at $\phi_\mathrm{det}$. We find that the SPI scenario requires $i_\star{\gtrsim}45^\circ$ for $\beta{<}30^\circ$. One possible geometry, shown in \autoref{fig:radio_period}, reproduces the observed emission with a visibility window of 3.7\,h.

\begin{table*}
\caption{Observational signatures distinguishing the three
scenarios for the origin of the ECMI emission from GJ\,725A.}
\label{tab:scenarios}
\centering
\renewcommand{\arraystretch}{1.3}
\begin{tabular}{lcccc}
\hline\hline
Diagnostic & Observation & Planetary aurora & Stellar activity &  Star--planet interaction \\
\hline
Periodicity & Radio & Planetary rot. ($\sim$hours) & None or $P_\mathrm{rot} \approx 103$\,d & $P_\mathrm{orb} \approx 11.2$\,d$^{(a)}$ \\
Coincidence with flares & Optical/X-ray &  ... & Yes & ... \\
Astrometry  & Radio VLBI & Offset up to $4''$ & Coincident & Coincident \\
New giant planet & PMA, RV & Yes & ... & ...  \\
Stellar $B$-field & ZDI & ... & Tens of gauss (\autoref{eq:escapefit}) & ${>}80$\,G  (\autoref{eq:subalfven},\autoref{fig:spi_energy}) \\
Wind base density $n_0$ & UV/X-ray lines & ... & \autoref{eq:escapefit} & ${<}2\times10^{8}\,\mathrm{cm}^{-3}$ (\autoref{eq:subalfven},\autoref{fig:spi_energy}) \\

\hline
\end{tabular}
\tablefoot{$^{(a)}$ Possibly modulated by the stellar rotation
period depending on the magnetic obliquity
\citep{2023MNRAS.524.6267K}.}
\end{table*}

\section{Discussion}\label{sec:discussion}

With a significance of 7.9$\sigma$ and a secure astrometric association, our detection represents the clearest case of coherent sub-100\,MHz emission from a system with a known planet to date. Previous candidate signals at these frequencies were either of marginal significance and/or lacked the angular resolution to identify an optical counterpart. At a distance of 3.53\,pc, it is furthermore among the nearest systems for which coherent radio emission has been observed, making it easily accessible for the follow-up campaigns discussed below.
 
While we identify ECMI as the likely emission mechanism, the available observational data do not yet allow us to establish the astrophysical origin of the power source. In the following, we  discuss the observational signatures that clarify the nature of the emission and the observations necessary for probing them.

Given that the available ultralow-frequency coverage of GJ\,725A is sparse, extended radio monitoring is the most direct way to understand its origin. A smoking-gun signature would be the detection of a periodicity. Emission with a period of 11.2\,d and concentrated near quadrature would be clear evidence of SPI with GJ\,725Ab, although the situation may differ if the system is spin-orbit misaligned \citep{2023MNRAS.524.6267K}. Detecting emission with a different, shorter period would point toward auroral emission of a gas planet (the Solar System giants have a rotational period of 10-17\,h). Emission due to stellar activity would either be stochastic or show some long-term periodicity with the stellar rotation ($\approx$103\,d) or activity cycle. The only existing telescopes capable of detecting $\approx100\,$mJy emission at frequencies $\leq50$\,MHz are NenuFAR \citep{zarka2020nenufar} and LOFAR, and only the latter is able to astrometrically associate the emission with one of the binary components. Pointed observations with the new LOFAR\,2.0 system will exhibit higher instantaneous sensitivity by a factor of 4 compared to the LoLSS data analyzed in this work \citep{2021A&A...652A..37E}.

LOFAR is also the only telescope capable of reaching sub-arc-second resolution at these frequencies by using very long baseline interferometry (VLBI). Although technically challenging, this could enable a precise localization of the radio emission. A signal due to a wide-orbit exoplanet may be up to $4''$ offset from the star (corresponding to the widest dynamically stable orbit), while emission in the stellar corona would have a negligible offset.

An important signature of radio emission due to stellar magnetic activity is a correlation between the radio emission and stellar flares at other wavelengths. By contrast, an exoplanetary origin would show no such correlation. To probe this signature, multiwavelength observations, combining radio monitoring with observations of stellar activity tracers such as H$\alpha$ and X-ray, are necessary. 
If the radio signal is due to auroral emissions from a giant planet on a wide orbit, improved searches in RV or PMA data, such as in the upcoming fourth data release of Gaia, will either reveal its presence or further constrain the permitted parameter space. 

Furthermore, the different scenarios are only viable for a limited part of the parameter space of our stellar wind and magnetic field model. For ECMI in the stellar magnetosphere, the formation criterion \autoref{eq:escapefit} must be fulfilled and thereby requires an $18-46\,\mathrm{G}$ base magnetic field for the coronal base densities considered.
The requirements for SPI are somewhat stricter. The stellar wind at the location of GJ\,725Ab must be sub-Alfv\'enic, and the planet must intercept sufficient Poynting flux to power the observed radio emission. This requires values of $B_{\star,0}$ ranging from ${\approx}80\,\mathrm{G}$ (for a tenuous wind) to $400\,\mathrm{G}$ (for a dense wind) for a conversion efficiency $\alpha_\mathrm{rad}{=}5\%$ and a planetary magnetic field of $B_\mathrm{p}{=}0.5\,$G.
It is possible to more accurately constrain our model parameters with new data. The most critical component is the strength and structure of the large-scale magnetic field of GJ\,725A, which can be probed using the Zeeman Doppler imaging (ZDI) method \citep[e.g.,][]{2021A&ARv..29....1K}.
Furthermore, the mass-loss rate and coronal base density can be observationally estimated, such as through Ly$\alpha$ absorption, which measures the hydrogen piled up in a star's astrosphere \citep{2021ApJ...915...37W}, and via density-sensitive coronal lines in the ultraviolet (UV) and X-ray
\citep{2009A&ARv..17..309G}. A high-resolution X-ray observation would further allow us to separate the flux of the two binary components, for which only the combined flux is currently known, thereby improving the accuracy of our estimate for the coronal temperature, $T$.
In \autoref{tab:scenarios}, we summarize the observational signatures and probes that can be used to distinguish between the three scenarios.

\section{Conclusions}\label{sec:conclusion}
While carrying out V-LoLSS, an untargeted circularly polarized radio survey with the LOFAR LBA, we discovered emission coincident with the solar-neighborhood M dwarf binary GJ\,725 at a distance of 3.53\,pc. Below, we summarize our observational findings:
\begin{enumerate}
    \item The source was detected at a significance of $7.9\sigma$ in a single 1\,h epoch on 11 December 2020, at frequencies below 50\,MHz, with a peak intensity $B_{I,46}=135\pm17\,\mathrm{mJy\,beam^{-1}}$ and is $74\pm16\%$ left-hand circularly polarized.
    \item The source is not detected above 50\,MHz, suggesting a strong spectral feature, such as a cutoff, in this range.
    \item Analyzing all other archival LOFAR observations covering the source does not yield any other detection.
    \item Based on an astrometric analysis of the source, we can clearly establish an association with the primary GJ\,725A.
\end{enumerate}
Our conclusions on the nature of the emission are as follows:
\begin{enumerate}[start=5]
    \item The inferred brightness temperature, $T_\mathrm{b,obs}{\sim}10^{14}\,$K, and degree of circular polarization far exceed the expectation for gyromagnetic and plasma emission. 
    \item We favor ECMI as the emission process, which can account for the high brightness temperature and circular polarization. The spectral cutoff at 50\,MHz implies a maximum  field strength $B_\mathrm{max}\approx 18\,$G in the emitting region.
    \item The only known exoplanet GJ\,725Ab is not massive enough to generate the required magnetic field. Thus, planetary auroral emission would need to originate from a yet-to-be-discovered massive planet. Existing constraints imply that such a planet would likely need to be in a wide orbit ($9000$--$29000\,$d).
    While the stellar wind cannot supply sufficient energetic electrons at such separations, an internal source such as a volcanically active satellite could. 
    \item The handedness of the emission is consistent with $x$-mode ECMI in a magnetic field with the orientation of the  disk-averaged longitudinal $B$-field of GJ\,725A.
    \item Visible ECMI emission can be generated from the stellar magnetosphere for a base wind density  $n_0 \lesssim 2.4\times10^8\left({B_\mathrm{\star,0}}/{50\,\mathrm{G}}\right)^{2}\,\mathrm{cm^{-3}}$. One possible mechanism for this are flaring coronal loops. While stellar activity indicators -- such as H$\alpha$ and X-ray luminosities, rotation periods, and optical flare rates -- indicate quiescence, an origin of the radio emission through a rare energetic flare remains a possibility. 
    \item Alternatively, SPI with the known close-in planet GJ\,725Ab could power the radio emission. This is consistent with the observed orbital phase of $0.24\pm0.05$ of the signal. For a sub-Alfv\'enic interaction to power the observed emission with $5\%$ radiative efficiency, a base magnetic field $B_{\star,0}>80\,$G is required.
    \end{enumerate}
Although the available data do not conclusively identify the origin of the emission, our finding constitutes both the lowest-frequency stellar detection and the first clear detection of coherent emission from a stellar system at ultralow frequencies. Future observations will be able to distinguish between the scenarios. 
We aim to carry out an extended ultralow-frequency monitoring campaign of the target in the context of the ExLOO and LLoCuSS LOFAR\,2.0 large programs. Repeated detections will allow us to distinguish between the scenarios based on the signal's periodicity.
Furthermore, a better understanding of the properties of the large-scale magnetic field of GJ\,725A via ZDI, the wind density via UV and X-ray spectroscopy, as well as simultaneous optical monitoring, will be critical for probing the feasibility of the different scenarios.
\begin{acknowledgements}
We thank the editor J.~A.~Caballero and the anonymous referee for their constructive comments on our manuscript, R.~D.~Kavanagh for advice on our model of the SPI maser, and S.~Bellotti for useful discussion on the magnetic field structure of GJ\,725A.
JRC acknowledges funding from the European Union via the European Research Council (ERC) grant Epaphus (project number 101166008) and from the Dutch research council (NWO) under the talent programme (Vidi grant VI.Vidi.243.136). HKV and CMC acknowledge funding from the European Research Council via the starting grant `STORMCHASER' (grant number 101042416). HKV acknowledges funding from  the Dutch research council under the talent programme (Vidi grant VI.Vidi.203.093). PZ acknowledges funding from the European Research Council (ERC) under the European Union’s Horizon 2020 research and innovation programme (grant agreement No 101020459 - Exoradio).
This publication used the Dutch SKA Regional Center prototype infrastructure provided by FuSE, the Fundamental Sciences e-Infrastructure, Dutch Research Council (NWO) Roadmap Project 184.035.004.
LOFAR is the Low Frequency Array designed and constructed by ASTRON. It has observing, data processing, and data storage facilities in several countries, which are owned by various parties (each with their own funding sources), and that are collectively operated by the ILT foundation under a joint scientific policy. The ILT resources have benefited from the following recent major funding sources: CNRS-INSU, Observatoire de Paris and Université d'Orléans, France; BMBF, MIWF-NRW, MPG, Germany; Science Foundation Ireland (SFI), Department of Business, Enterprise and Innovation (DBEI), Ireland; NWO, The Netherlands; The Science and Technology Facilities Council, UK; Ministry of Science and Higher Education, Poland; The Istituto Nazionale di Astrofisica (INAF), Italy. This research made use of the Dutch national e-infrastructure with support of the SURF Cooperative (e-infra 180169) and the LOFAR e-infra group. The Jülich LOFAR Long Term Archive and the German LOFAR network are both coordinated and operated by the Jülich Supercomputing Centre (JSC), and computing resources on the supercomputer JUWELS at JSC were provided by the Gauss Centre for Supercomputing e.V. (grant CHTB00) through the John von Neumann Institute for Computing (NIC).
This research made use of the University of Hertfordshire high-performance computing facility and the LOFAR-UK computing facility located at the University of Hertfordshire and supported by STFC [ST/P000096/1], and of the Italian LOFAR IT computing infrastructure supported and operated by INAF, and by the Physics Department of Turin university (under an agreement with Consorzio Interuniversitario per la Fisica Spaziale) at the C3S Supercomputing Centre, Italy. The authors acknowledge the use of Claude Opus (Anthropic) to assist with copyediting and improving the clarity of individual paragraphs. All outputs were reviewed, verified, and finalized by the authors.
\end{acknowledgements}
\bibliographystyle{aa}
\bibliography{references}
\appendix
\section{Raw dynamic spectra}\label{sec:app_dynspec}
\begin{figure}[!h]
    \centering
        \includegraphics[width=0.9\linewidth]{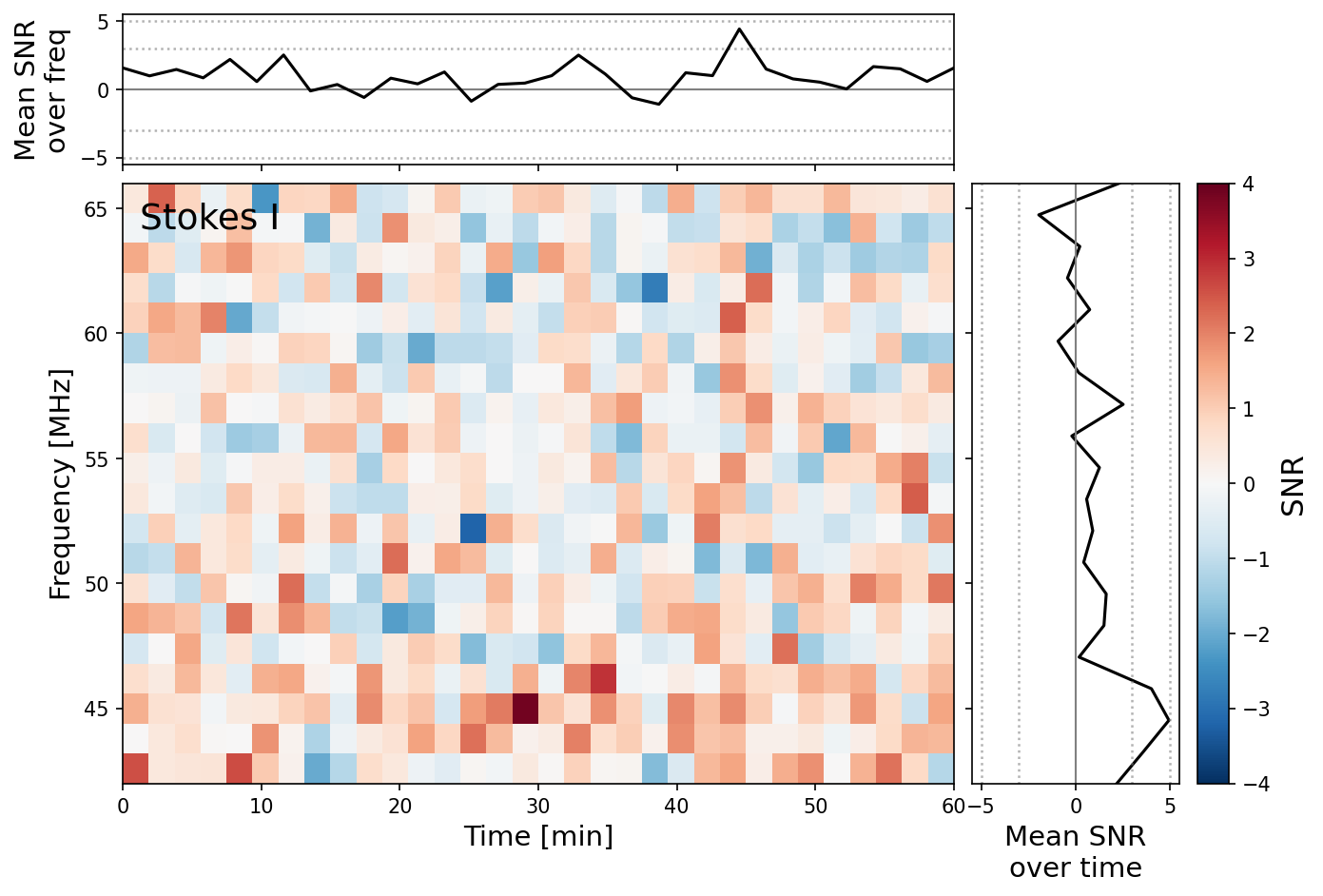}
        \includegraphics[width=0.9\linewidth]{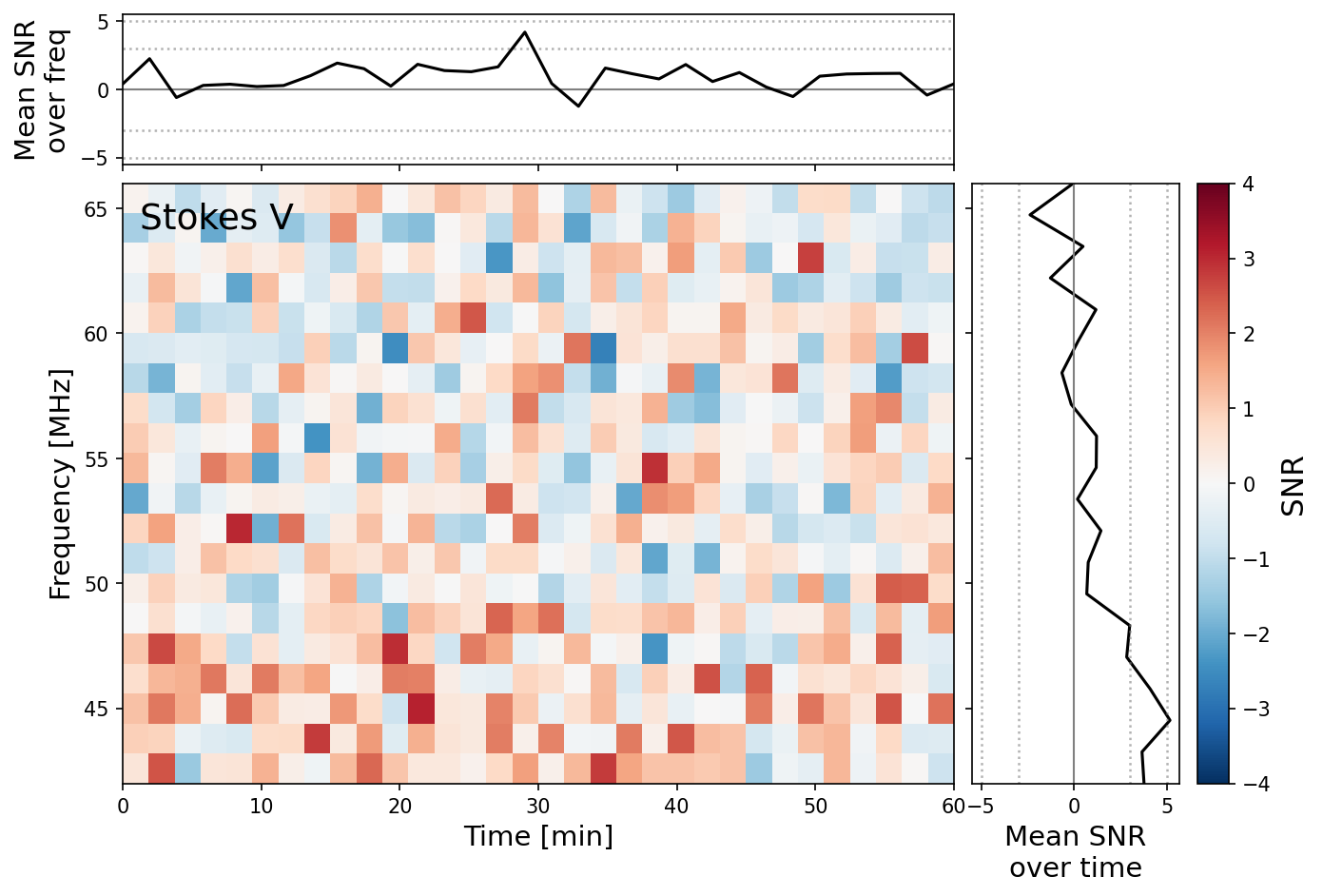}
    \caption{Dynamic spectra as in \autoref{fig:dynspecs} but showing the S/N in bins of 1.17\,MHz and 112\,s.}
    \label{fig:dynspecs_raw}
\end{figure}

\FloatBarrier
For reference, the unsmoothed dynamic spectra, in bins of 1.17\,MHz and 112\,s, are shown in \autoref{fig:dynspecs_raw}.
\FloatBarrier
\section{Wind and magnetic field model}\label{sec:app_wind}
We modeled the stellar wind of GJ\,725A using the model of \citet{1958ApJ...128..664P} as implemented by \citet{2026A&A...714A.401D}, which uses as input parameters the stellar radius and mass, the coronal temperature $T_0$, the base magnetic field $B_{\star,0}$, and wind base density $n_{0}$.
We assumed an isothermal stellar wind, with a temperature $T_0=1.8\times10^{6}$\,K that we derived from the X-ray data following the relation of \citet{2015A&A...578A.129J}. The ROSAT X-ray measurement of GJ\,725 does not resolve the binary components. Thus, we assumed that both contribute equally to the total luminosity of $L_\mathrm{X}=7.7\times10^{26}\,\mathrm{erg\,s^{-1}}$ \citep{2004A&A...417..651S}.

\subsection{Coronal base density}
The available data on GJ\,725A do not allow for a measurement of $n_0$. Therefore, we employed the scaling relation derived by \citet{2021ApJ...915...37W} (see also \citealt{2024ApJ...976...65R}) from measurements of other M dwarfs to estimate $n_0$ from the mass-loss rate,
\begin{equation}
{\dot{M}_\star}
=
\dot{M}_\odot\frac{{R_\star^2}}{R_\odot^2}\left(\frac{F_{X,\star}}{F_{X,\odot}}\right)^{0.77} 
\approx 3.5\times10^{-15}\,\mathrm{M_\odot\,yr^{-1}},
\end{equation} 
where $\dot{M}_\odot=2\times10^{-14}\,\mathrm{M_\odot\,yr^{-1}}$ and $F_{X,\odot}=3.2\times10^{4}\,\mathrm{erg\,s^{-1}\,cm^{-2}}$ are the solar mass-loss rate and X-ray surface flux, respectively. 
We then obtained the base number density of the stellar corona using our wind model solution for the base velocity $v_0=7.6\,\mathrm{km\,s^{-1}}$ as
\begin{equation}
    n_0=\frac{\dot{M}_\star}{\mu_\mathrm{e}{m_\mathrm{H}}v_04\pi{R}_\star^2},
\end{equation}
where $\mu_\mathrm{e}=1.15$ is the mean molecular weight per electron (corresponding to a mean molecular mass $\mu=0.6$ considering the total particle population), and $m_\mathrm{H}$ is the hydrogen mass.
Using our estimate of $L_\mathrm{X}$ and the wind model, we obtained $n_0=2.0\times10^{7}\,\mathrm{cm^{-3}}$. This estimate carries significant uncertainty, given its indirect inference. Thus, we also explored values an order of magnitude lower and higher in the parameter space of $n_0$.

\begin{figure}
    \centering
    \includegraphics[width=0.9\linewidth]{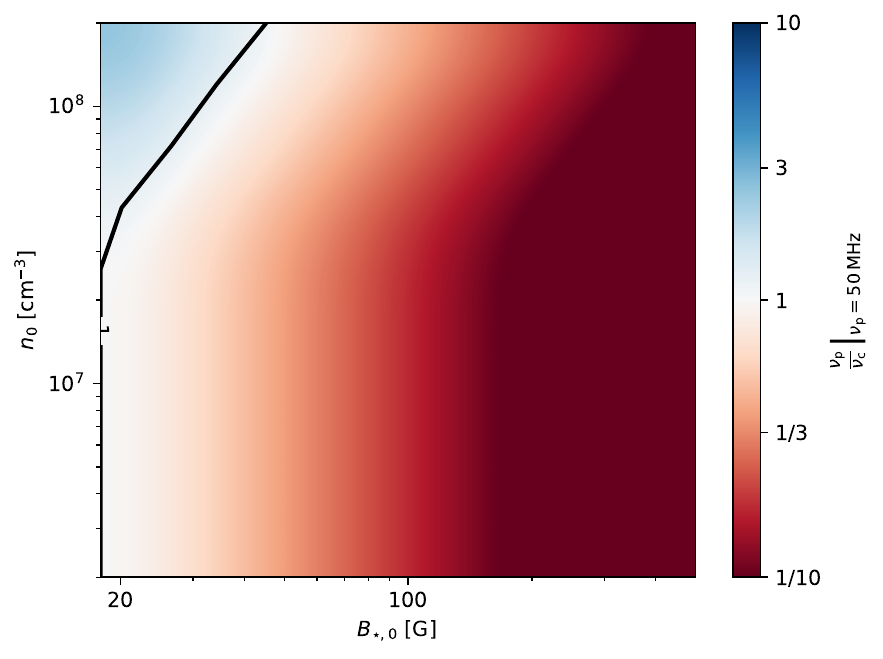}
    \includegraphics[width=0.9\linewidth]{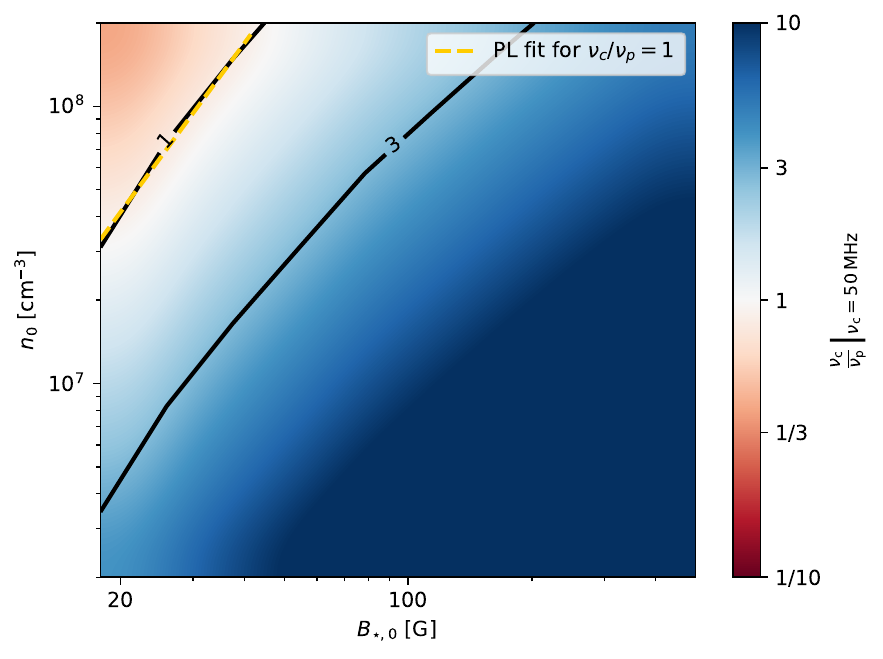}
    \includegraphics[width=0.9\linewidth]{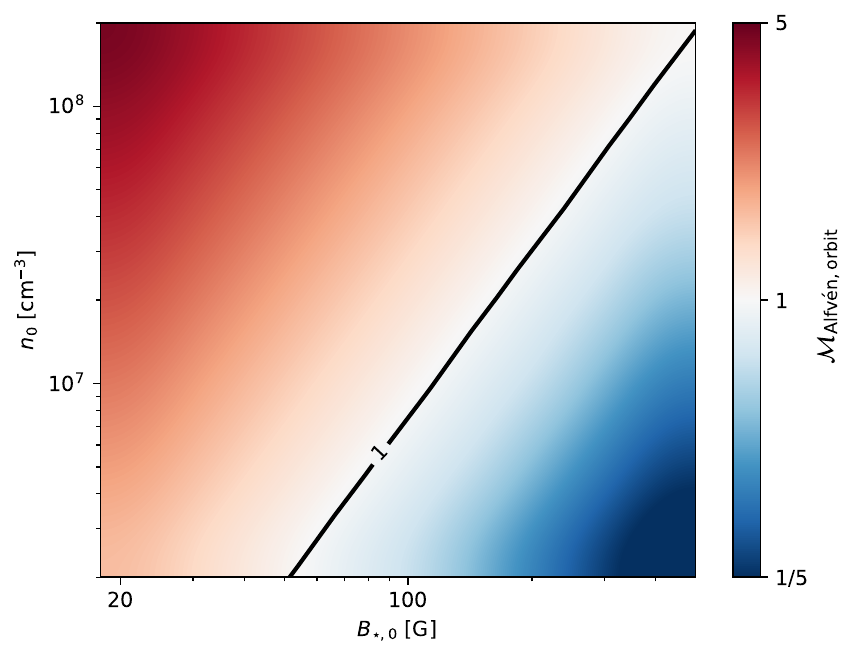}
    \caption{Top: Ratio between  plasma and cyclotron frequency at the location where the plasma frequency equals the maximum observed frequency of 50\,MHz. This represents a formation condition for plasma emission. 
    Center: Ratio between cyclotron and plasma frequency at the location where the cyclotron frequency equals 50\,MHz. This represents the formation condition for ECMI. The yellow line illustrates the fit to the points where the ratio is unity shown in \autoref{eq:escapefit}. 
    Bottom: Alfv\'enic Mach number at the orbital radius of GJ\,725Ab.}
    \label{fig:conditions}
\end{figure}

\subsection{Magnetic field}
We assumed the large-scale stellar magnetic field close to the star is a rotation-aligned dipole, $B(r) = B_{\star,0} \left(R_\star / r\right)^{3}$, consistent with the predominantly poloidal and axisymmetric field topology recovered for GJ\,725\,A by \citet{Donati2023}. Since the surface field strength relevant for the global wind structure is only weakly constrained by the available measurements - it has an upper bound in the mean unsigned surface field of $\langle B \rangle \sim 490\pm70\,$G \citep{Cristofari2023}, we treated $B_0$ as a free parameter and explored the range $18 \leq B_0 \leq 500\,$G. The lower limit is the minimum magnetic field necessary for ECMI at 50\,MHz.
Our magnetic field model includes a transition from the closed-field to the open-field regime ($B\propto{r}^{-2}$) at a distance $R_\mathrm{open}$, which we took to be the radius at which the magnetic pressure equals the sum of the thermal and ram pressure of the stellar wind. 

From the wind and field model, the local Alfv\'en velocity $v_\mathrm{A}(r) = B(r)/\sqrt{4\pi\rho(r)}$ and the Alfv\'enic Mach number $M_\mathrm{A} = v_\mathrm{rel}/v_\mathrm{A}$ follow directly, where $v_\mathrm{rel}$ is the wind velocity in the rest
frame of the planet, including its orbital motion. We caution that this model neglects magnetocentrifugal acceleration, non-isothermal heating, and the interplay between field topology and wind expansion.
Depending on the parameters, the radius where the cyclotron frequency is equal to the maximum observed frequency of our signal is 0--2\,R$_\star$ above the stellar surface. 

\section{TESS flare rate}\label{sec:app_flarerate}
GJ\,725 was extensively monitored by TESS in sectors 14--26, 40, 41, 47--57, 59, 60, and 73--86, spanning July 2019 to December 2024. As discussed by \citet{Cortes-Zuleta2025}, the angular separation of the binary ($\approx11''$) is smaller than the
TESS pixel scale of $21''$, so that depending on the sector both components fall either in the same or in adjacent pixels, and the flux of the two stars cannot be separated reliably. The rate derived here is therefore an upper limit for GJ\,725A, as it may include flares from the secondary.

We retrieved the 2-min cadence data using \texttt{lightkurve} \citep{2018ascl.soft12013L} and identified flares with \texttt{stella} \citep{2020JOSS....5.2347F}, a convolutional neural network that assigns a flare probability $p \in [0,1]$ to each cadence. We defined candidate flares as events with at least five contiguous cadences at $p > 0.5$, corresponding to a minimum duration of $10\,$min. This yielded 19 candidates, of which we rejected 8 upon visual inspection of the light curves as instrumental systematics (sectors 20, 23, and 82). The remaining flares correspond to an upper limit on the flare rate of GJ\,725A of $0.013\,\mathrm{d^{-1}}$. 

GJ\,725 was not observed by TESS at the epoch of our LOFAR detection. However, no flares at all are detected in the sectors preceding December 2023, with ten events occurring after April 2024 at a rate of $<0.05\,\mathrm{d^{-1}}$. The optical flare rate in the years bracketing our detection is therefore consistent with zero, providing no indication of enhanced flaring activity at that time.

\section{SPI maser model}\label{sec:app_maser}
\begin{figure}
    \centering
    \includegraphics[width=0.99\linewidth]{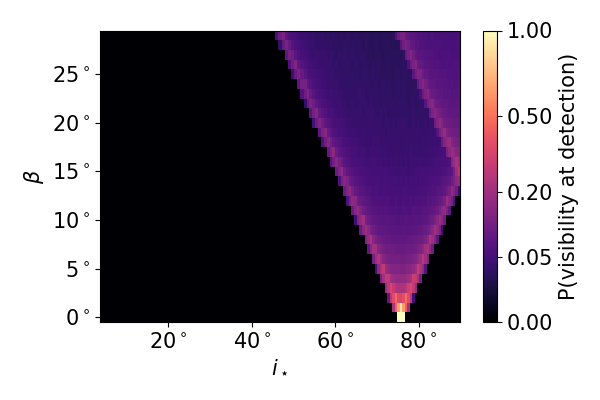}
    \caption{Probability of visible SPI maser emission at the orbital phase of the radio detection ($\phi_\mathrm{det}=0.24\pm0.05$) as a function of stellar inclination and magnetic obliquity.}
    \label{fig:geometry}
\end{figure}

We explored possible geometries of the SPI scenario based on the \texttt{MASER} code\footnote{\url{https://github.com/robkavanagh/maser/tree/main}} \citep{2023MNRAS.524.6267K}. 
The available observational data on GJ\,725A are not sufficient to closely constrain the geometry; however, we attempted a comparison with a plausible first-order model. 
For this, we assumed a dipolar stellar surface magnetic field of 100\,G (this choice has only a minor effect on the geometry), with an inclination of the planetary orbit of 70$^\circ$ (aligned with the binary orbit as suggested by \citealt{Cortes-Zuleta2025}). We further assumed an opening angle of the hollow ECMI emission cone of 75$^\circ$, with a thickness of 1.5$^\circ$ \citep{1998JGR...10326649Q,zarka2004jupiter}. The parameters that are largely unconstrained are the stellar inclination $i_\star$ and the magnetic obliquity $\beta$. We explored the probability of visible maser emission at the orbital phase of our detection of $\phi_\mathrm{det}=0.24\pm0.05$ on a uniformly sampled grid of $\beta\in[0^\circ,30^\circ]$ and $\cos{i_\star}\in[0,1]$, by simulating the maser cone geometry across the full range of initial stellar rotation phases.  In \autoref{fig:geometry}, we display the parameter space and the corresponding visibility probability. 
The maximum probability is reached for an aligned dipole ($\beta=0$) and a stellar inclination of $i_\star\approx76.4^\circ$, with these parameters, the maser is always visible at  $\phi_\mathrm{det}$, with a visibility window of 3.7\,h. 
For the considered parameter space, the maser is only visible for inclinations $i_\mathrm{\star,max\,prob.}>45^\circ$. The lower the magnetic obliquity, the closer the inclination needs to be to $i_\mathrm{\star,max\,prob.}$ for the maser to be visible at the phase of detection.
In \autoref{fig:radio_period}, we display the visibility windows for $i_\mathrm{\star,max\,prob.}$  and $\beta=0$. This precise scenario is in apparent tension with the non-detections around $\phi=0.77$. However, this can be alleviated already with a small shift in the orbital phase ($\Delta\phi\approx0.004$, corresponding to $<0.1\sigma_\mathrm{conj}$), well below the uncertainty on $\phi_\mathrm{det}$. To showcase a possible scenario that is in agreement with the observations, we added this small shift to the displayed visibility windows.
A more constrained modeling of the geometry will be possible once a ZDI map of GJ\,725A becomes available, or with additional radio detections.

\end{document}